%% file: MADBench_arXiv.tex
\documentclass[letterpaper]{article} 
\usepackage[preprint]{aaai2027}  

\usepackage[hyphens]{url}  
\usepackage{graphicx} 
\usepackage{natbib}  
\usepackage{caption} 
\usepackage{algorithm}
\usepackage{algorithmic}
\usepackage{soul, xcolor}

\usepackage{newfloat}
\usepackage{listings}
\DeclareCaptionStyle{ruled}{labelfont=normalfont,labelsep=colon,strut=off} 
\floatstyle{ruled}
\newfloat{listing}{tb}{lst}{}
\floatname{listing}{Listing}

\usepackage{booktabs}

\usepackage{amssymb}
\usepackage{amsmath}
\usepackage{multirow}
\usepackage{tabularx}

\title{MADBench: A Benchmark for \underline{M}odality-Aware \underline{A}udio \underline{D}eepfake Detection}
\newcommand{\sysname}{MADBench}

\author{
Yanqiu Li\textsuperscript{\rm 1},
Yang Xiao\textsuperscript{\rm 1},
Jisheng Bai\textsuperscript{\rm 2},
Bin Chen\textsuperscript{\rm 1},
Hong Jia\textsuperscript{\rm 3},
Ting Dang\textsuperscript{\rm 1}
}

\affiliations{
\textsuperscript{\rm 1}The University of Melbourne, Melbourne, Australia\\
\textsuperscript{\rm 2}Xi'an University of Posts and Telecommunications, Xi'an, China\\
\textsuperscript{\rm 3}The University of Auckland, Auckland, New Zealand
}

\begin{document}

\maketitle

\input{00-abs}


\input{01-intro}

\input{02-related}

\section{Dataset Construction}

\input{03-design}

\subsection{Source Preprocessing and Splitting}

We build \sysname{} from AVSpeech~\cite{R48}, which comprises
real-world YouTube videos with visible speakers and naturally
paired speech. For each source clip, the original video stream
is retained, while MossFormer2~\cite{R49} estimates a speech
stem \(\hat{s}\) from the mixed waveform \(x\). We define the
non-speech residual as \(\hat{e}=x-\hat{s}\), following the
mixture-consistency principle that the separated components
should reconstruct the original mixture~\cite{R64}. After
quality control, \(\hat{s}\) and \(\hat{e}\) serve as the
source-derived real speech and environmental-audio components.

We validate the decomposition using independent speech-content
and leakage checks. All speech stems
are successfully transcribed with \(95.9\%\) mean ASR coverage,
while independent VAD and ASR checks detect no speech in any
environmental residual. All residuals also satisfy a minimum
environmental-energy criterion. These checks ensure that the
retained clips contain intelligible speech and perceptible
environmental audio without detectable speech leakage.

Before fake-component generation, source clips are assessed for
benchmark eligibility. We retain clips with valid media, usable
English transcripts, sufficient spoken content, perceptible
environmental audio, and no substantial music or media
playback. The retained clips are approximately \(4\)--\(12\)
seconds long, making them suitable for controlled audio
replacement and final audio-visual assembly. To reduce speaker
and source leakage, all clips from the same detected speaker
cluster or related source identifiers are assigned to the same
training, validation, or test split.

After source-level quality control, 1,892 clips remain eligible
for subsequent four-way benchmark construction. These clips are
partitioned approximately \(7{:}1{:}2\) into training,
validation, and test sets. The split is fixed before
fake-component generation, and all derived samples inherit the
split assignment of their source clip.

\input{04-fakespeech}

\input{05-fakeenvaudio}

\subsection{Sample Assembly and Dataset Statistics}

\begin{table}[t]
\centering
\scriptsize
\setlength{\tabcolsep}{2.5pt}
\renewcommand{\arraystretch}{1.06}

\begin{tabular*}{\columnwidth}{
@{\extracolsep{\fill}}
l
r
}
\toprule
\textbf{Statistic} & \textbf{Count} \\
\midrule

\multicolumn{2}{l}{\textit{Component Construction}} \\

Generation-eligible source clips
& \(\mathbf{1{,}892}\) \\

Speech components (generated / QC-retained)
& \(34{,}056 / 24{,}614\) \\

Environmental components (generated / QC-retained)
& \(22{,}987 / 21{,}388\) \\

\midrule

\multicolumn{2}{l}{\textit{\sysname{} Dataset}} \\

Source clips (train / val. / test)
& \(\mathbf{1{,}378}\;(986 / 143 / 249)\) \\




Samples per protocol (train / val. / test)
& \(\mathbf{5{,}512}\;(3{,}944 / 572 / 996)\) \\

Final audio-visual samples (across 3 protocols)
& \(\mathbf{16{,}536}\) \\

\bottomrule
\end{tabular*}

\caption{Statistics of the \sysname{} dataset.}
\label{tab:dataset_statistics}
\end{table}

Final assembly combines the QC-passed speech and environmental components into complete audio-visual samples under a shared media construction policy. For each source clip, four samples are constructed: \textbf{a reassembled real sample, a speech-only fake, an environment-only fake, and a joint fake}. All four variants use the same real video. Within each source, the selected fake speech component is shared by the
speech-only and joint-fake variants, while the selected fake
environmental component is shared by the environment-only and
joint-fake variants. This ensures that controlled comparisons
differ in only one audio component.
Importantly, the real sample is also reassembled from separated real speech and real environmental audio through the same mixing procedure as the fake samples, ensuring that all sample types share an identical construction process, so that labels reflect component differences rather than assembly differences. 

Component selection is balanced across generation settings to prevent any single generator or branch from dominating the benchmark. Speech components are distributed across the six speech generation settings. Environmental components are balanced across TTA, VTA, ATA branches and generation models. Final audio is standardized and mixed using source-relative RMS normalization, so that speech and environmental-audio levels remain tied to the original source audio.
The final audio is then paired with the real source video to produce the audio-visual sample. Table~\ref{tab:dataset_statistics} summarizes the core statistics of \sysname{}.

\input{06-eval}

\input{07-results}

\section{Conclusion}

We introduced \sysname{}, a component-level audio-visual deepfake benchmark, where speech and environmental audio are independently manipulated over authentic video. Experiments show that existing pretrained A-V detectors transfer poorly and zero-shot omni models remain unreliable, whereas frozen A-V encoders support strong detection and attribution. Environmental manipulation is generally easier to detect and can obscure speech-specific cues, while video contributes mainly to scene-consistency reasoning rather than manipulation detection. These findings highlight the importance of component-aware audio-visual deepfake evaluation.

\bibliography{aaai2027}

\end{document}

%% file: 00-abs.tex
\begin{abstract}
Recent advances in speech synthesis and audio generation have made high-fidelity acoustic forgery low-cost and difficult to attribute, enabling a realistic attack scenario in which speech and background audio are independently manipulated over otherwise authentic video. Yet existing research either focuses on visual manipulation, addresses speech detection in isolation, or conflates speech and non-speech audio as a single undifferentiated audio stream, overlooking the distinct forensic challenges posed by background audio. This conflation is consequential: the two acoustic components arise from fundamentally different generative mechanisms, exhibit distinct artifact profiles, and pose different challenges to detection systems. We introduce \textbf{\sysname{}}, the first benchmark that treats speech and environmental audio as distinct acoustic components, enabling component-aware evaluation of audio deepfake detection across independently manipulated forgery sources. We benchmark representative state-of-the-art detectors and multimodal large language models under a unified protocol. Our experiments reveal that environmental audio manipulation is more detectable than synthetic speech across general-purpose encoders, while existing pretrained detectors fail on both acoustic components, and manipulated environmental audio asymmetrically degrades speech deepfake detection, findings entirely invisible under the single-label paradigm of prior benchmarks. \sysname{} establishes a rigorous foundation for future research into robust, component-aware audio deepfake detection.
\end{abstract}

%% file: 01-intro.tex
\section{Introduction}
The rapid advancement of artificial intelligence has altered the threat landscape of synthetic media. Deepfakes, AI-manipulated content designed to deceive, have evolved from face-swap techniques into sophisticated, multimodal fabrications capable of fooling both human perception and automated detection systems alike~\cite{R2,R3}. While significant research effort has been devoted to detecting visual forgeries~\cite{R5}, a quieter and arguably more insidious manipulation vector has received far less scrutiny: the replacement or synthesis of audio in otherwise authentic video.

Consider a realistic and increasingly common attack scenario. An adversary obtains a genuine video recording, a politician's speech, a corporate announcement, a personal conversation, and leaves the visual content entirely intact. Instead, they replace or synthesize the speech to alter what is being said~\cite{R50,R51}, and manipulate the background audio to make the fabrication acoustically coherent and convincing~\cite{R14,R37}. The resulting deepfake is visually authentic by construction, yet entirely deceptive in meaning. This setting is operationally attractive precisely because visual forgery detection has matured considerably~\cite{R6,R60}, making audio-domain manipulation a lower-cost, lower-risk avenue for bad actors.

This attack surface exposes a critical and underappreciated distinction that the research community has largely failed to formalize: 
\emph{speech and environmental audio are distinct acoustic components}.
Speech carries linguistic and paralinguistic content, what is said and how it is said, and is generated by fundamentally different synthesis pipelines than non-speech audio~\cite{R8,R9}. Background soundscapes, ambient acoustics, environmental noise, and foley-style effects constitute a separate generative and perceptual domain, one that plays an equally important role in establishing the perceived authenticity of a video. A fabricated crowd reaction, a synthetically generated room ambiance, or an artificially composed acoustic scene can all serve to reinforce a deceptive narrative~\cite{R17,R17-1,R17-2,R17-3}, yet none of these are captured by speech-centric detection models.

Despite the growing importance of audio-only deepfakes, \textbf{no existing benchmark is designed for authentic-video scenarios with independently manipulated speech and non-speech audio}. Existing datasets either focus on visual manipulations, treat speech as representative of the entire acoustic stream~\cite{R1,R4}, or collapse all audio manipulations into a single label~\cite{R12}. As a result, the research community currently lacks the tools to answer even basic questions: \emph{How detectable are synthesized speech tracks when the video is genuine? How do non-speech audio artifacts differ from speech artifacts under detection? Does the presence of manipulated background audio confound or assist speech deepfake detectors?} Without benchmarks that explicitly disentangle these acoustic components, it is impossible to systematically evaluate detector robustness or understand the limitations of current approaches.

To address this gap, we introduce the first benchmark designed specifically for audio-component deepfake detection in authentic-video settings, where speech and non-speech audio are manipulated independently while the visual content remains unchanged. Unlike existing datasets, our benchmark explicitly annotates these two acoustic sources as separate forgery components, enabling fine-grained evaluation of component-specific detection performance as well as their interactions. This formulation provides a more realistic and challenging testbed for developing next-generation audiovisual deepfake detectors. Extensive experiments reveal a striking finding: 
although audio-visual encoders consistently detect environmental audio manipulation more reliably than speech, existing pretrained detectors fail on both acoustic components, and manipulated environmental audio degrades speech deepfake detection without the converse holding, a cross-component effect that remains invisible when speech and non-speech audio are collapsed into a single label, leading to an incomplete assessment of detector robustness.

Therefore, our contributions are summarized as follows:
\begin{itemize}

    \item \textbf{A new benchmark.} We introduce \sysname{}, the first benchmark for audio-component deepfake detection in authentic-video
    settings, establishing a realistic evaluation task where speech and non-speech audio are manipulated independently.

    \item \textbf{A modality-aware dataset.} We present a dataset with independently synthesized and annotated speech and non-speech audio, enabling fine-grained evaluation of acoustic deepfakes beyond speech-centric settings.

    \item \textbf{A comprehensive testbed.} We benchmark representative state-of-the-art deepfake detection models under a unified evaluation protocol, providing strong baselines for future research.
    
\end{itemize}

%% file: 02-related.tex
\section{Related Work}

\subsection{Audio-Visual Deepfake Benchmarks}

Early benchmarks such as FaceForensics++ and DeeperForensics~\cite{R1,R4} established face-forgery detection as the dominant paradigm, but their evaluation targets remain visual or binary real/fake.  Recent audio-visual benchmarks extend the task beyond visual-only forgery. FakeAVCeleb \cite{R5} combines fake faces with synthesized or cloned lip-synced audio, while LAV-DF \cite{R6} and AV-Deepfake1M \cite{R7} introduce content-driven audio-visual manipulation and temporal localization. These benchmarks are important for multimodal fake detection, but they generally treat audio manipulation as speech, identity, or content manipulation rather than separating speech and environmental audio as independent components. 

\subsection{Speech and Environmental Audio Deepfakes}

Speech-only spoofing benchmarks~\cite{R10,yi2023add,R13} have supported research on detecting synthetic and converted speech, replay attacks, and partially spoofed utterances.  However, these benchmarks focus on speech and do not evaluate non-speech background audio. Recent advances in text-to-audio generation have made background-sound manipulation increasingly realistic, creating a forensic challenge that speech-centric benchmarks are not designed to address. This has given rise to a nascent line of work on environmental audio deepfake detection: EnvSDD~\cite{R17} and Compspoof~\cite{zhang2026compspoof} focus on audio-only environmental sound deepfake detection, while VCapAV~\cite{R18} studies audio-visual environmental manipulation. However, EnvSDD and Compspoof provide no visual grounding, while VCapAV does not independently control speech and environmental audio as separate forgery components. Neither benchmark supports the systematic evaluation of cross-component interactions. MADBench addresses these gaps by jointly controlling speech manipulation, environmental audio manipulation, and scene consistency within a single component-level audio-visual benchmark. 

\subsection{Deepfake Detection Models}

Deepfake detectors use different cues depending on modality. Speech anti-spoofing models~\cite{R34} such as AASIST~\cite{R11} focus on synthetic or converted speech artefacts, while audio-visual methods such as AVForensics~\cite{R31} compare cross-modal signals through lip-sync, speech-mouth alignment, or multimodal fusion.  However, none of these methods can attribute a manipulation to a specific acoustic component, nor can they be directly evaluated on environmental audio deepfakes that MADBench is designed to expose.  Beyond task-specific detectors, pretrained audio and multimodal encoders such as CLAP~\cite{R24} and ImageBind~\cite{R27} provide general-purpose representations that serve as non-task-specific baselines for probing whether component-level deepfake signals are recoverable without forensic fine-tuning. Recent omni models such as MiniCPM-o~\cite{R45} also represent a qualitatively different detection paradigm: rather than relying on learned forensic features, they apply large-scale multimodal pretraining to reason about audio-visual content zero-shot. Evaluating these models on MADBench directly tests whether general multimodal intelligence transfers to component-level forensic detection, a question that existing benchmarks cannot address due to their lack of component-level annotation.

%% file: 03-design.tex
\subsection{Overall Design Principles}

\begin{figure*}[t]
    \centering
    \includegraphics[width=\textwidth]{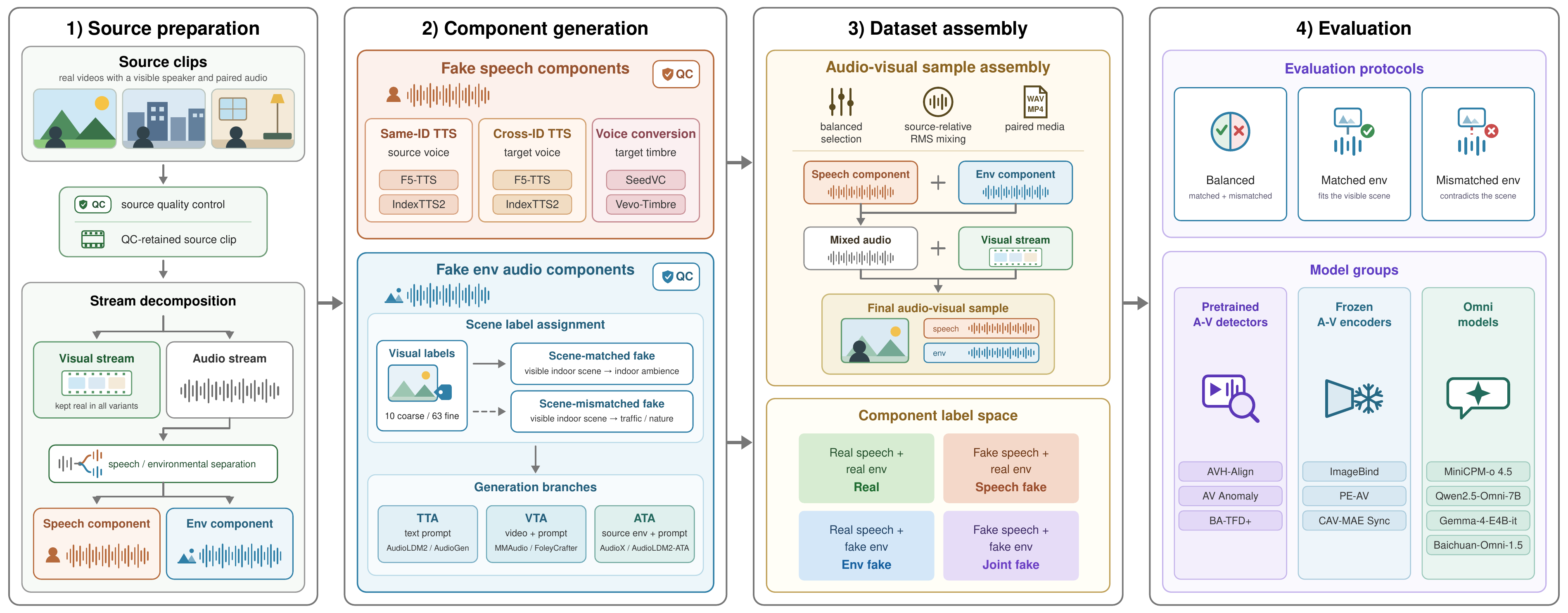}
    \caption{Overview of the \sysname{} dataset construction and benchmark evaluation pipeline.}
    \label{fig:caves_avdf_pipeline}
\end{figure*}

As illustrated in Figure~1, \sysname{} follows a
modality-component hierarchy: each video comprises audio and visual modalities, while
the audio modality is further decomposed into speech and
environmental audio components. The visual stream is held fixed across all variants of a given source clip, while manipulation is applied exclusively to the speech component, the environmental audio component, or both. This design choice ensures that any difference in model behavior across variants is attributable to audio manipulation alone, rather than being confounded by visual artifacts.

To further control evaluation difficulty and probe distinct model capabilities, we introduce a \emph{scene-consistency axis} orthogonal to the manipulation type. Manipulated samples are categorized as either \emph{scene-matched}, where the substituted environmental audio remains semantically plausible given the visible scene, or \emph{scene-mismatched}, where it is intentionally inconsistent with the visual context. This distinction is critical because it separates two fundamentally different detection strategies: recognizing low-level synthesis artifacts or component-level manipulation cues, versus reasoning about high-level semantic coherence between the audio and visual streams. Together, these two axes, manipulation component and scene consistency form the structural foundation of the benchmark.

%% file: 04-fakespeech.tex
\subsection{Fake Speech Generation}
Fake speech is generated to match the duration of the source real speech and placed into the original speech regions, with non-speech regions left silent. We use three speech synthesis techniques to ensure broad coverage of diverse speaker manipulation strategies.

\paragraph{Speech synthesis.} We first use \emph{same-identity Text-to-Speech (TTS) models}, which synthesises the source transcript using the source speaker as reference, preserving apparent speaker identity. F5-TTS \cite{R50} and IndexTTS2 \cite{R51} are used for this branch. 
\emph{Cross-identity TTS} synthesises the same transcript with a different target speaker. Voice conversion transforms the source speech toward a target speaker's timbre while preserving source content and timing. SeedVC \cite{R52} and Vevo-Timbre \cite{R53} are used for this branch. Together, the three branches yield six generation settings, reducing dependence on any single manipulation mechanism.

\paragraph{Speaker pairing.} Cross-identity generation follows a fixed source-target pairing table, constrained to same-gender pairs from different source videos and different detected speaker clusters. Speaker embeddings are used to exclude near-duplicate voices, and target-speaker usage is balanced to prevent a small number of identities from dominating the pool. Each source clip is paired with four target speakers, providing redundancy for QC-based selection. ECAPA \cite{R54} similarity between source and target speakers is low (mean = 0.0268, max = 0.3040), confirming acoustic distinctiveness.

\paragraph{Quality control.} Generated speech components are assessed against automatic QC
criteria covering generation duration, timeline alignment and ASR-based transcript preservation. Furthermore, same-identity TTS is checked for source-speaker preservation; cross-identity TTS and voice conversion are checked for target-speaker similarity and source-speaker leakage.

%% file: 05-fakeenvaudio.tex
\subsection{Fake Environmental Audio Generation}
Environmental-audio generation produces full-duration background
components to replace the real environmental track. To construct
scene-matched and scene-mismatched variants systematically, we
require a shared definition of what sounds belong to each type of
visible scene. We therefore build a visual-scene taxonomy to
guide both audio generation and mismatch selection. The pipeline
comprises three stages: \emph{taxonomy construction, audio
generation, and QC}.

\paragraph{Taxonomy Construction. }
To ensure reproducible scene matching, we first define a taxonomy of visual scenes, where each label specifies (i) expected environmental sounds, (ii) excluded sound categories, and (iii) mismatch scene types.

The taxonomy is constructed exclusively from the training and validation set.
Qwen2.5-VL-32B \cite{R61} summarizes the visual scene, expected environmental sounds, and potential audio confounders (e.g., speech or music), providing standardized descriptions for taxonomy construction.

The resulting taxonomy is refined through an LLM-assisted review followed by manual verification to remove redundant labels, identify missing scene categories, and validate mismatch relationships. The final taxonomy contains 10 coarse and 63 fine-grained scene categories spanning indoor, outdoor, transportation, industrial, educational, and natural environments. Since the environmental component represents background ambience only, all labels explicitly exclude speech, singing, music, and media playback.

After the taxonomy is finalized, all source clips are automatically assigned a fine-grained scene label
following the predefined taxonomy. These labels determine prompt construction, scene-matched generation, and valid mismatch selection throughout the pipeline.

\paragraph{Environmental Audio Generation.}
We generate environmental audio using three complementary generation paradigms. The \emph{Text-to-Audio (TTA)} branch synthesizes background audio from textual scene descriptions using AudioLDM2-TTA \cite{R16} and AudioGen \cite{R14}. The \emph{Video-to-Audio (VTA)} branch conditions generation on both visual content and text prompts using MMAudio \cite{R56} and FoleyCrafter \cite{R57}. The \emph{Audio-to-Audio (ATA)} branch edits or regenerates the environmental component while conditioning on the source background audio using AudioX \cite{R55} and AudioLDM2-ATA \cite{R16}. Together, these branches span prompt-conditioned, video-conditioned, and source-audio-conditioned generation, reducing dependence on any single generation strategy.

For scene-matched generation, all branches produce environmental audio consistent with the labeled visual scene. For scene-mismatched generation, the TTA branch directly synthesizes audio from prompts describing a different scene category. Since VTA and ATA are inherently conditioned on the source clip, mismatched samples are instead created by replacing the generated environmental track with one originating from a semantically incompatible scene. Replacement tracks are selected within the same data split while excluding the same source clip and related source groups, and reuse is limited to reduce identity leakage and shortcut learning.

All prompts are generated automatically from the scene taxonomy using fixed templates that incorporate coarse and fine-grained scene labels, inferred environmental sound events, candidate mismatch scenes, and clip duration. This rule-based procedure ensures consistent prompt quality and enforces the exclusion of speech and music across all generated samples.

\paragraph{Quality Control.}
Generated components are retained if they satisfy QC criteria. Checks cover duration alignment, severe silence or clipping, detectable speech or music content, and valid replacement pairing for mismatch variants. Voice activity detection (VAD) and ASR detect speech leakage; AST-based event checks and CLAP-style audio-prompt matching scores serve as diagnostic signals.

%% file: 06-eval.tex
\section{Evaluation Setup}

\subsection{Tasks and Protocols}

All samples in \sysname{} share an authentic visual stream and differ only
in the states of their two audio components. We denote the
four sample types as \(R\) (real speech and real environmental
audio), \(S\) (manipulated speech and real environmental
audio), \(E\) (real speech and manipulated environmental
audio), and \(Q\) (both audio components manipulated).

We evaluate three core tasks. Binary any-fake detection
separates \(R\) from the three manipulated states
\(S\), \(E\), and \(Q\). Four-way classification directly
predicts \(R/S/E/Q\). Component-level evaluation independently
determines whether the speech and environmental audio have
been manipulated, allowing both components to be identified
as fake in a \(Q\) sample.

The three tasks are evaluated under three core protocols.
The \emph{Balanced} protocol includes both scene-matched and
scene-mismatched environmental manipulations. The
\emph{Matched} protocol contains only manipulated
environmental audio that remains plausible for the visible
scene, whereas the \emph{Mismatched} protocol uses
scene-incompatible environmental audio. All three core protocols
retain the same manipulation labels and evaluate audio
manipulation detection and attribution, supplemented by
controlled analysis protocols for scene consistency, generator
generalization, and input modality ablation.

We report standard metrics, including ROC-AUC, EER, accuracy,
and Macro-F1. Task-specific diagnostics are defined in the
corresponding result subsections.

\subsection{Model Groups and Evaluation Procedure}

We evaluate three complementary model groups to examine
whether component-level audio forgery cues arise from
task-specific forensic training, general-purpose audio-visual
pretraining, or zero-shot multimodal reasoning.

First, to assess the transferability of existing forensic
systems, we evaluate the pretrained A-V
deepfake detectors AVH-Align~\cite{R32}, AV Anomaly~\cite{R63}, and BA-TFD+~\cite{R60}.
Their native scores are used for binary direct transfer, while
lightweight prediction heads are fitted to frozen
detector outputs for tasks not supported by the original models.

Second, to determine whether general-purpose A-V
representations contain recoverable manipulation and
cross-modal correspondence signals, we evaluate ImageBind ~\cite{R27},
PE-AV Base ~\cite{peav}, and CAV-MAE Sync ~\cite{araujo2025cav} as frozen representation
encoders with lightweight task-specific heads. Audio-only
encoders \cite{R24, R25, R26}, and pretrained audio detectors
\cite{R11, xiao2024xlsrmamba, kulkarni2026compactsslbackbonesmatter, huang2026audiomosaic} are included as
unimodal reference baselines.

Third, to test whether large-scale multimodal pretraining
produces zero-shot manipulation detection, component
attribution, and scene-consistency reasoning, we evaluate
Qwen2.5-Omni-7B~\cite{R44}, MiniCPM-o~4.5~\cite{R45}, Gemma-4-E4B-it~\cite{team2026gemma}, and
Baichuan-Omni-1.5~\cite{li2025baichuan} using fixed task-specific prompts without 
additional training.

All pretrained detectors and encoders remain frozen. 
The prediction heads use the same
logistic-regression architecture but are trained independently
for each task on the training split, with thresholds calibrated
on the validation split. Omni models are evaluated zero-shot using
fixed task-specific prompts without additional training. The
task-specific prediction targets, notation, and prompt
assignments are introduced in the corresponding result setups.

%% file: 07-results.tex
\section{Results and Analysis}
\label{sec:results_analysis}

We organise our analysis around four diagnostic questions:
(1) whether existing models can detect and identify the
manipulations;
(2) whether models respond 
differently to speech and environmental audio 
manipulation; (3) whether models leverage visual-audio 
inconsistency or rely purely on acoustic artifacts;
and (4) whether visual input improves 
detection beyond audio alone.

\subsection{Can Existing Models Detect Deepfakes?}

\begin{table}[t]
\centering
\scriptsize
\setlength{\tabcolsep}{1.1pt}
\renewcommand{\arraystretch}{1.08}

\begin{tabular*}{\columnwidth}{
@{\extracolsep{\fill}}
l
c c c c
}
\toprule
\textbf{Model} &
\textbf{Balanced} &
\textbf{Matched} &
\textbf{Mismatched} &
\textbf{Mean} \\
\midrule

\multicolumn{5}{l}{
\textbf{(a) Binary Any-Fake Detection}
\textbf{ (AUC$\uparrow$ / EER$\downarrow$)}
} \\

\multicolumn{5}{l}{
\textit{Pretrained A-V Detectors (direct transfer)}
} \\

AVH-Align
& \textbf{0.525} / 0.491
& \textbf{0.532 / 0.487}
& 0.515 / 0.498
& 0.524 / 0.492 \\

AV Anomaly
& 0.518 / \textbf{0.481}
& 0.512 / 0.492
& 0.524 / 0.481
& 0.518 / \textbf{0.484} \\

BA-TFD+
& 0.524 / 0.487
& 0.526 / 0.493
& \textbf{0.552 / 0.475}
& \textbf{0.534} / 0.485 \\




\midrule

\multicolumn{5}{l}{
\textit{Frozen A-V Encoders (with \(H_{\mathrm{bin}}\))}
} \\

ImageBind A--V
& 0.912 / 0.163
& 0.912 / 0.160
& 0.918 / 0.165
& 0.914 / 0.162 \\

PE-AV Base
& 0.943 / \textbf{0.120}
& 0.930 / \textbf{0.131}
& 0.953 / \textbf{0.108}
& 0.942 / \textbf{0.120} \\

CAV-MAE Sync
& \textbf{0.954} / 0.123
& \textbf{0.949} / 0.133
& \textbf{0.958} / 0.112
& \textbf{0.954} / 0.122 \\

\midrule

\multicolumn{5}{l}{
\textit{Omni Models (zero-shot \(P_{\mathrm{bin}}\))}
} \\

\shortstack[l]{MiniCPM-o 4.5}
& \textbf{0.621 / 0.418}
& \textbf{0.600 / 0.430}
& \textbf{0.634 / 0.410}
& \textbf{0.618 / 0.419} \\

\shortstack[l]{Qwen2.5-Omni-7B}
& 0.540 / 0.479
& 0.523 / 0.490
& 0.542 / 0.470
& 0.535 / 0.480 \\

\shortstack[l]{Gemma-4-E4B-it}
& 0.557 / 0.449
& 0.558 / 0.444
& 0.560 / 0.450
& 0.558 / 0.448 \\

\shortstack[l]{Baichuan-Omni-1.5}
& 0.553 / 0.456
& 0.465 / 0.523
& 0.462 / 0.523
& 0.494 / 0.501 \\

\midrule

\multicolumn{5}{l}{
\textbf{(b) Four-Way Manipulation Classification}
\textbf{ (Acc.$\uparrow$ / Macro-F1$\uparrow$)}
} \\

\multicolumn{5}{l}{
\textit{Pretrained A-V Detectors (with \(H_{\mathrm{4way}}\))}
} \\

AVH-Align
& 0.271 / 0.201
& 0.268 / 0.192
& 0.268 / 0.199
& 0.269 / 0.197 \\

AV Anomaly
& 0.269 / 0.187
& 0.254 / 0.177
& 0.265 / 0.188
& 0.263 / 0.184 \\

BA-TFD+
& \textbf{0.286 / 0.240}
& \textbf{0.295 / 0.257}
& \textbf{0.295 / 0.245}
& \textbf{0.292 / 0.248} \\

\midrule

\multicolumn{5}{l}{
\textit{Frozen A-V Encoders (with \(H_{\mathrm{4way}}\))}
} \\

ImageBind A--V
& 0.590 / 0.588
& 0.594 / 0.593
& 0.597 / 0.595
& 0.594 / 0.592 \\

PE-AV Base
& \textbf{0.745 / 0.745}
& \textbf{0.711 / 0.709}
& \textbf{0.756 / 0.756}
& \textbf{0.737 / 0.737} \\

CAV-MAE Sync
& 0.711 / 0.708
& 0.681 / 0.676
& 0.696 / 0.696
& 0.696 / 0.693 \\

\midrule

\multicolumn{5}{l}{
\textit{Omni Models (zero-shot \(P_{\mathrm{4way}}\))}
} \\

\shortstack[l]{MiniCPM-o 4.5}
& 0.249 / 0.106
& 0.250 / 0.105
& \textbf{0.256} / 0.118
& 0.252 / 0.110 \\

\shortstack[l]{Qwen2.5-Omni-7B}
& \textbf{0.250 / 0.146}
& \textbf{0.257 / 0.156}
& 0.252 / \textbf{0.150}
& \textbf{0.253 / 0.151} \\

\shortstack[l]{Gemma-4-E4B-it}
& \textbf{0.250} / 0.100
& 0.250 / 0.100
& 0.250 / 0.100
& 0.250 / 0.100 \\

\shortstack[l]{Baichuan-Omni-1.5}
& \textbf{0.250} / 0.117
& 0.254 / 0.125
& 0.250 / 0.117
& 0.251 / 0.120 \\
\bottomrule
\end{tabular*}

\caption{
(a) Binary any-fake detection and 
(b) four-way \(R/S/E/Q\) classification
across the three core protocols.
Mean denotes the average across protocols.
}
\label{tab:overall_av_detection}
\end{table}

\paragraph{Setup.}

We compare the three model groups at two levels of difficulty:
binary detection tests whether they distinguish authentic from
manipulated audio, while four-way classification tests whether
the learned cues support fine-grained manipulation attribution.
The binary head \(H_{\mathrm{bin}}\) distinguishes fully real samples \(R\) from
\(S/E/Q\), whereas \(H_{\mathrm{4way}}\) directly predicts \(R/S/E/Q\).
Pretrained A-V detectors are evaluated using their native predictions to measure direct transfer. For 
four-way classification, detector features are frozen 
and a logistic-regression head \(H_{\mathrm{4way}}\) is 
fitted on top. Frozen A-V encoders are similarly adapted with a binary and 4-way logistic-regression head.
Omni models use the zero-shot Binary Detection Prompt
(\(P_{\mathrm{bin}}\)) and Four-Way Classification Prompt
(\(P_{\mathrm{4way}}\)) respectively.

\paragraph{Pretrained A-V detectors fail to transfer.}
The three pretrained detectors perform near chance in binary direct transfer (mean AUC: 0.518–0.534). To test whether this stems from a distribution shift between their training data and \sysname{}, we train a benchmark-specific binary classification head on their frozen features. Performance remains essentially unchanged, with only a 0.012 AUC gain for BA-TFD+, indicating that the frozen representations themselves lack transferable cues rather than being limited by the original decision heads. A likely reason is that existing A-V detectors are designed for face-centric videos, whereas \sysname{} contains diverse natural scenes beyond a single face. 
Four-way classification is likewise close to chance. The matched and mismatched protocols yield comparable performance across both tasks, suggesting that scene mismatch has little impact on either detection or manipulation identification.

\paragraph{Frozen A-V encoders transfer strongly.}
In clear contrast, all three frozen encoders achieve over 0.91 mean AUC for binary classification and maintain strong performance on four-way classification. Compared with task-specific deepfake detectors, these broadly pretrained audio-visual encoders provide substantially more transferable representations. Performance consistently improves under the mismatched protocol, indicating that audio-visual scene inconsistency provides an additional discriminative signal. However, the improvement is modest across all encoders, suggesting that matched scenes already contain sufficient information for reliable detection. 

\paragraph{Omni models detect coarse anomalies but not components.}
Zero-shot omni models show limited binary sensitivity, 
with MiniCPM-o~4.5 achieving the best mean AUC of 0.618, 
but this does not extend to fine-grained four-way classification with all models 
remaining near chance. This gap reflects a fundamental mismatch between general multimodal pretraining and the forensic reasoning required to attribute manipulation to a specific acoustic source. Notably, performance does not improve under the mismatched protocol, confirming that visible scene-audio contradiction provides no additional attribution signal for these models.

\paragraph{Takeaway.}
The three model groups establish a clear transfer hierarchy: pretrained A-V detectors fail to capture the required forgery structure, zero-shot omni models provide only coarse sensitivity, and frozen A-V encoders support both reliable detection and fine-grained manipulation attribution.

\subsection{Do Models Respond Differently to Speech and Environmental Audio Manipulations?}

\begin{table}[t]
\centering
\scriptsize
\setlength{\tabcolsep}{1.0pt}
\renewcommand{\arraystretch}{1.07}

\begin{tabular*}{\columnwidth}{
@{\extracolsep{\fill}}
l
c
c c c c
c
c
}
\toprule
\textbf{Model} &
\textbf{Comp.} &
\multicolumn{4}{c}{\textbf{AUC$\uparrow$}} &
\(\boldsymbol{\Delta_{\mathrm{E-S}}}\) &
\(\boldsymbol{G_{\mathrm S},\,G_{\mathrm E}}\) \\
\cmidrule(lr){3-6}

&
&
\textbf{Bal.} &
\textbf{Mat.} &
\textbf{Mis.} &
\textbf{Mean} &
&
\\
\midrule

\multicolumn{8}{l}{
\textit{Pretrained A-V Detectors (with \(H_{\mathrm S}\) and \(H_{\mathrm E}\))}
} \\

\multirow{2}{*}{AVH-Align}
& S
& \textbf{0.550}
& \textbf{0.555}
& \textbf{0.535}
& \textbf{0.547}
& \multirow{2}{*}{-0.034}
& 0.008 \\

& E
& 0.517
& 0.508
& 0.513
& 0.513
&
& -0.010 \\

\addlinespace[1pt]

\multirow{2}{*}{AV Anomaly}
& S
& 0.516
& 0.510
& 0.506
& 0.511
& \multirow{2}{*}{0.033}
& 0.012 \\

& E
& 0.549
& 0.533
& 0.547
& 0.543
&
& -0.012 \\

\addlinespace[1pt]

\multirow{2}{*}{BA-TFD+}
& S
& 0.484
& 0.509
& 0.500
& 0.498
& \multirow{2}{*}{0.094}
& 0.030 \\

& E
& \textbf{0.583}
& \textbf{0.594}
& \textbf{0.599}
& \textbf{0.592}
&
& -0.047 \\

\midrule

\multicolumn{8}{l}{
\textit{Frozen A-V Encoders (with \(H_{\mathrm S}\) and \(H_{\mathrm E}\))}
} \\

\multirow{2}{*}{ImageBind A-V}
& S
& 0.797
& 0.803
& 0.793
& 0.798
& \multirow{2}{*}{0.106}
& 0.138 \\

& E
& 0.898
& 0.896
& 0.919
& 0.904
&
& 0.002 \\

\addlinespace[1pt]

\multirow{2}{*}{PE-AV Base}
& S
& \textbf{0.895}
& \textbf{0.886}
& \textbf{0.896}
& \textbf{0.892}
& \multirow{2}{*}{0.044}
& 0.046 \\

& E
& 0.940
& 0.914
& 0.954
& 0.936
&
& -0.012 \\

\addlinespace[1pt]

\multirow{2}{*}{CAV-MAE Sync}
& S
& 0.857
& 0.847
& 0.844
& 0.849
& \multirow{2}{*}{0.113}
& 0.137 \\

& E
& \textbf{0.963}
& \textbf{0.951}
& \textbf{0.974}
& \textbf{0.963}
&
& -0.017 \\

\midrule

\multicolumn{8}{l}{
\textit{Omni Models (zero-shot \(P_{\mathrm{comp}}\))}
} \\

\multirow{2}{*}{
\shortstack[l]{MiniCPM-o 4.5}}
& S
& 0.470
& 0.463
& 0.489
& 0.474
& \multirow{2}{*}{0.022}
& 0.036 \\

& E
& \textbf{0.499}
& 0.486
& \textbf{0.504}
& \textbf{0.496}
&
& 0.031 \\

\addlinespace[1pt]

\multirow{2}{*}{\shortstack[l]{Qwen2.5-Omni-7B}}
& S
& \textbf{0.500}
& \textbf{0.502}
& 0.492
& \textbf{0.498}
& \multirow{2}{*}{-0.007}
& 0.008 \\

& E
& 0.489
& \textbf{0.497}
& 0.487
& 0.491
&
& 0.006 \\

\multirow{2}{*}{\shortstack[l]{Gemma-4-E4B-it}}
& S
& 0.487
& 0.479
& 0.474
& 0.480
& \multirow{2}{*}{-0.005}
& 0.034 \\

& E
& 0.477
& 0.470
& 0.479
& 0.475
&
& 0.034 \\

\multirow{2}{*}{\shortstack[l]{Baichuan-Omni-1.5}}
& S
& 0.486
& 0.485
& \textbf{0.494}
& 0.488
& \multirow{2}{*}{+0.000}
& -0.006 \\

& E
& 0.497
& 0.483
& 0.485
& 0.488
&
& -0.003 \\

\bottomrule
\end{tabular*}

\caption{
Component-level binary detection performance for speech (S) and
environmental-audio (E) manipulation across the balanced (Bal.),
scene-matched (Mat.), and scene-mismatched (Mis.) protocols.
\(G_{\mathrm{S}}\) and \(G_{\mathrm{E}}\) denote the
speech and environmental interference gaps reported in the S and E rows,
respectively.
}
\label{tab:component_level_analysis}
\end{table}

\paragraph{Setup.}

Although speech has dominated prior audio deepfake detection research, it remains unclear whether this produces a systematic advantage of speech deepfake detection over environmental audio deepfake detection. To test this, we decompose the four-way task into two binary detection problems: a speech detection head 
\(H_{\mathrm{S}}\) that distinguishes samples with fake speech (\(S, Q\)) from those without (\(R, E\)), and an environmental detection head \(H_{\mathrm{E}}\) that distinguishes samples with fake environmental audio (\(E, Q\)) from those without (\(R, S\)). The two heads 
are fitted independently on frozen detector outputs, whereas omni models use the zero-shot component detection prompt \(P_{\mathrm{comp}}\). We summarize the relative detection advantage as 
\(\Delta_{\mathrm{E-S}} = \overline{\mathrm{AUC}}_{\mathrm{E}} 
- \overline{\mathrm{AUC}}_{\mathrm{S}}\), where a 
positive value indicates stronger sensitivity to 
environmental manipulation.

We further measure component interference: whether detecting one manipulated component becomes harder when the other component is also manipulated. For speech, 
\(G_{\mathrm{S}} = \mathrm{AUC}(S\text{ vs. }R) - 
\mathrm{AUC}(Q\text{ vs. }E)\) compares speech detection 
when environmental audio is real versus fake. For 
environmental audio, \(G_{\mathrm{E}} = 
\mathrm{AUC}(E\text{ vs. }R) - \mathrm{AUC}(Q\text{ 
vs. }S)\) compares environmental detection when speech 
is real versus fake. A positive gap indicates that 
manipulating the other component degrades detection of 
the target component.

\paragraph{Transferred detectors show no stable component preference.}
The pretrained A-V detectors do not exhibit the expected
universal speech advantage. The comparable performance across balanced, matched, and mismatched cases across E and S confirms that there are no differences in detecting speech and audio deepfakes using existing detectors. This may still be due to the misalignment of the training objectives of these detectors, which are more face-centric and fails in both speech and environmental audio settings.

\paragraph{Environmental manipulation is consistently easier for frozen A-V encoders.}
Environmental manipulation is detected more accurately than speech manipulation across all encoders and evaluation protocols. Notably, the environmental advantage persists under the matched protocol, indicating that it cannot be explained solely by audio-visual scene inconsistency. Instead, the consistent gains across all three encoders suggest that large-scale audio-visual pretraining yields transferable representations that are particularly effective for environmental audio manipulation, with PE-AV's more diverse pretraining leading to the most balanced performance across the two components.

\paragraph{Omni models do not separate the two components.}
Zero-shot omni models remain close to chance for both speech
and environmental manipulation, with no consistent direction
in \(\Delta_{\mathrm{E-S}}\). Their general multimodal
perception capabilities therefore do
not translate into component-specific forensic decisions
without task supervision.

\paragraph{Fake environmental audio obscures speech-specific deepfake detection.}

\(G_{\mathrm S}\) and \(G_{\mathrm E}\) are close to
zero for most of the models, indicating limited cross-component interference. The clear exceptions are ImageBind and CAV-MAE Sync,
whose \(G_{\mathrm S}\) values reach \(0.138\) and \(0.137\),
respectively. The positive \(G_{\mathrm S}\) shows that fake environmental audio obscures 
speech-specific cues; however, environmental detection remains
stable when speech is also manipulated.
This reflects the
component structure, as environmental
manipulation spans the acoustic context of the clip, while
speech manipulation is concentrated in speech-active regions
and must be isolated from the surrounding soundscape.

\paragraph{Takeaway.}
Environmental manipulation is consistently
easier to detect than speech and also the dominant source of cross-component interference. 
Models pretrained on diverse sound events and audio-visual correspondence
perform better than speech-focused detectors and zero-shot omni
models.

\subsection{Do Models Exploit Scene-Audio Consistency?}

\begin{table}[t]
\centering
\scriptsize
\setlength{\tabcolsep}{1.0pt}
\renewcommand{\arraystretch}{1.06}


\resizebox{\columnwidth}{!}{%
\begin{tabular}{
l
ccc
ccc
ccc
}
\toprule
&
\multicolumn{3}{c}{\textbf{D-R}} &
\multicolumn{3}{c}{\textbf{S-R}} &
\multicolumn{3}{c}{\textbf{D-S}} \\
\cmidrule(lr){2-4}
\cmidrule(lr){5-7}
\cmidrule(lr){8-10}

\textbf{Model} &
\(\mathbf{P}\!\uparrow\) &
\(\mathbf{Sh}\!\uparrow\) &
\(\boldsymbol{\Delta_{\mathrm{pair}}}\!\uparrow\) &
\(\mathbf{P}\!\uparrow\) &
\(\mathbf{Sh}\!\uparrow\) &
\(\boldsymbol{\Delta_{\mathrm{pair}}}\!\uparrow\) &
\(\mathbf{P}\!\uparrow\) &
\(\mathbf{Sh}\!\uparrow\) &
\(\boldsymbol{\Delta_{\mathrm{pair}}}\!\uparrow\) \\
\midrule

\multicolumn{10}{l}{
\textit{Pretrained A-V Detectors (with \(H_{\mathrm{SC}}^{\mathrm{FM}}\))}
} \\

AVH-Align
& 0.500 & 0.502 & -0.002
& 0.501 & 0.503 & -0.002
& 0.500 & 0.505 & -0.005 \\

AV Anomaly
& 0.499 & 0.500 & -0.001
& 0.491 & 0.492 & -0.002
& 0.497 & 0.498 & -0.001 \\

BA-TFD+
& \textbf{0.525} & \textbf{0.524} & \textbf{+0.001}
& \textbf{0.507} & \textbf{0.509} & -0.002
& \textbf{0.523} & \textbf{0.519} & \textbf{+0.004} \\

\midrule




\multicolumn{10}{l}{
\textit{Frozen A-V Encoders (with \(H_{\mathrm{SC}}^{\mathrm{FM}}\))}
} \\

ImageBind
& 0.744 & 0.719 & \textbf{+0.025}
& 0.705 & 0.728 & -0.023
& \textbf{0.551} & 0.531 & \textbf{+0.020} \\

PE-AV Base
& 0.726 & 0.731 & -0.005
& 0.728 & 0.712 & \textbf{+0.016}
& 0.532 & \textbf{0.544} & -0.012 \\

CAV-MAE Sync
& \textbf{0.808} & \textbf{0.795} & +0.013
& \textbf{0.789} & \textbf{0.782} & +0.008
& 0.523 & 0.507 & +0.016 \\

\midrule

\multicolumn{10}{l}{
\textit{Omni Models (zero-shot \(P_{\mathrm{SC}}\))}
} \\

Qwen2.5-Omni-7B
& 0.500
& 0.490
& \textbf{+0.010}
& 0.502
& 0.490
& \textbf{+0.012}
& 0.495
& 0.498
& -0.004 \\

MiniCPM-o 4.5
& \textbf{0.510}
& \textbf{0.525}
& -0.014
& 0.489
& 0.484
& +0.005
& 0.484
& \textbf{0.520}
& -0.035 \\

Gemma-4-E4B-it
& 0.502
& 0.498
& +0.004
& 0.500
& 0.497
& +0.002
& 0.505
& 0.505
& +0.000 \\

Baichuan-Omni-1.5
& 0.501
& 0.524
& -0.024
& \textbf{0.505}
& \textbf{0.516}
& -0.010
& \textbf{0.507}
& 0.493
& \textbf{+0.015} \\

\bottomrule
\end{tabular}%
}

\caption{
Audio-visual scene-consistency performance (AUC) with paired
(\(P\)) and shuffled (\(Sh\)) video inputs.
\(\Delta_{\mathrm{pair}}\) measures the benefit of the intended video
and only positive gains are highlighted.
}
\label{tab:scene_audio_correspondence}
\end{table}

\paragraph{Setup.}
When a model succeeds in the fixed-video setting, is
it responding to acoustic cues in the audio, or to the semantic
mismatch between what is seen and heard? A model driven by
acoustic cues can make its decision without using the video,
whereas detecting scene-audio mismatch requires exploiting
the cross-modal relationship between vision and sound.

To separate these two sources of evidence, we construct three
controlled scene-consistency conditions. \emph{D-R} uses real
environmental audio from different coarse scenes; \emph{S-R} uses
real audio from the same coarse scene but different fine-grained
labels; and \emph{D-S} compares scene-consistent and
scene-inconsistent samples in which both environmental tracks
are synthetic.
Comparing D-R with S-R tests whether models
move beyond coarse-category matching to fine-grained scene
consistency, while comparing D-R with D-S examines how
synthetic acoustic characteristics affect consistency recognition.

We formulate scene consistency as a binary task, where each sample is classified as scene-consistent or scene-inconsistent.
Since this target differs from the authenticity and component labels defined above, supervised models 
use a dedicated pooled Full-Mix Scene-Consistency Head
\(H_{\mathrm{SC}}^{\mathrm{FM}}\). The head is trained on the union of the paired
D-R, S-R, and D-S samples with both speech and
environmental audio retained, rather than fitting a separate head for each
condition. Omni models make the same binary decision using the
zero-shot Scene-Consistency Prompt (\(P_{\mathrm{SC}}\)).
At test time, the Paired condition (\(P\)) uses the intended
video, whereas the Shuffled condition (\(Sh\)) keeps the audio
and label unchanged and replaces only the video. We measure the contribution of the correct visual stream as
\(
\Delta_{\mathrm{pair}}
=
\mathrm{AUC}_{P}
-
\mathrm{AUC}_{Sh},
\)
where a positive value indicates that the intended video improves scene-consistency detection.

\paragraph{Pretrained detectors do not capture scene consistency.}
All three pretrained
detectors remain close to chance across D-R, S-R, and D-S.
Their paired and shuffled results are nearly identical,
indicating that the correct video contributes little to the
decision. BA-TFD+ shows only a small improvement over chance on
D-R and D-S, and this advantage largely remains after video
shuffling, suggesting that it is driven by audio-side cues
rather than scene grounding. Their objectives focus on speech-visual alignment rather than
the consistency between environmental audio and the visual
scene, limiting their transfer to environmental
scene-consistency reasoning.

\paragraph{Frozen encoders use both audio cues and visual context.}
Frozen A-V encoders perform strongly on both D-R and S-R.
CAV-MAE Sync reaches paired AUCs of \(0.808\) and \(0.789\),
respectively, while PE-AV maintains nearly identical
performance across the two conditions. 
ImageBind shows the largest drop, from \(0.744\) to \(0.705\),
indicating comparatively weaker fine-grained consistency recognition. 
Overall, the strong S-R results indicate that these
representations capture fine-grained scene correspondence rather
than relying only on coarse scene categories.

Frozen A-V encoders exhibit the clearest paired-video benefit
among the three model groups, showing that they make the most
effective use of visual scene information.
These gains are most apparent under D-R and D-S, whereas S-R yields
smaller and model-dependent benefits, reflecting the greater
difficulty of fine-grained correspondence. The high shuffled
scores on D-R and S-R further show that audio-internal cues also
remain an important source of evidence.

\paragraph{Fine-grained alignment provides the most consistent visual benefit.}
Under D-S, models across all three groups perform close to
chance. Because both consistency labels contain synthetic
environmental audio, synthesis status and generic generation
artifacts cannot directly determine the label. This broad
performance drop exposes the difficulty of scene grounding when
real-recording consistency cues are unavailable. Nevertheless,
CAV-MAE Sync achieves positive \(\Delta_{\mathrm{pair}}\) 
in all three conditions, making it the
only encoder that consistently benefits from the correct video.
This suggests that fine-grained audio-frame alignment preserves
sample-specific scene correspondence more effectively than
relying only on global cross-modal similarity.

\paragraph{Omni models fail at zero-shot scene grounding.}
All four omni models remain close to chance across the three
conditions, and their paired-shuffled differences are small and
inconsistent. Their responses therefore provide no consistent evidence of
zero-shot visual grounding.

\paragraph{Takeaway.}
Pretrained detectors and zero-shot
omni models remain near chance with negligible or inconsistent
video pairing gains, whereas frozen A-V encoders capture both
fine-grained consistency cues and modest visual grounding.
The broad performance drop on D-S identifies
synthetic scene correspondence as the most challenging setting.

\subsection{Is Audio Alone Sufficient for Manipulation Detection?}

\begin{table}[t]
\centering
\scriptsize
\setlength{\tabcolsep}{1.2pt}
\renewcommand{\arraystretch}{1.07}

\begin{tabular*}{\columnwidth}{
@{\extracolsep{\fill}}
l
c
c c c c
}
\toprule
\textbf{Model} &
\textbf{Input} &
\shortstack{\textbf{Binary}\\\textbf{AUC$\uparrow$}} &
\shortstack{\textbf{4-way}\\\textbf{Macro-F1$\uparrow$}} &
\shortstack{\textbf{Speech}\\\textbf{AUC$\uparrow$}} &
\shortstack{\textbf{Env.}\\\textbf{AUC$\uparrow$}} \\
\midrule
\multicolumn{6}{l}{
\textit{Frozen Audio Encoders}
} \\
CLAP
& A
& \textbf{0.969}
& \textbf{0.775}
& \textbf{0.915}
& \textbf{0.966} \\
BEATs
& A
& \underline{0.926}
& \underline{0.659}
& 0.815
& \underline{0.958} \\
WavLM
& A
& 0.904
& 0.623
& \underline{0.862}
& 0.858 \\
\midrule
\multicolumn{6}{l}{
\textit{Pretrained Audio Detectors}
} \\
AASIST
& A
& 0.766
& 0.497
& 0.725
& 0.845 \\
XLSR-Mamba
& A
& 0.857
& 0.627
& 0.840
& 0.874 \\
DF-Arena-1B
& A
& \textbf{0.967}
& \textbf{0.826}
& \textbf{0.991}
& \textbf{0.945} \\
AudioMosaic
& A
& \underline{0.949}
& \underline{0.732}
& \underline{0.898}
& \underline{0.935} \\
\midrule
\multicolumn{6}{l}{
\textit{Frozen A-V Encoders}
} \\
\multirow{2}{*}{ImageBind}
& A
& 0.961
& 0.705
& 0.890
& 0.950 \\
& A-V
& 0.914
& 0.592
& 0.798
& 0.904 \\
\addlinespace[1pt]
\multirow{2}{*}{PE-AV Base}
& A
& \underline{0.963}
& \textbf{0.787}
& \textbf{0.937}
& \underline{0.968} \\
& A-V
& 0.942
& 0.737
& 0.892
& 0.936 \\
\addlinespace[1pt]
\multirow{2}{*}{CAV-MAE Sync}
& A
& \textbf{0.975}
& \underline{0.763}
& \underline{0.904}
& \textbf{0.985} \\
& A-V
& 0.954
& 0.693
& 0.849
& 0.963 \\
\midrule
\multicolumn{6}{l}{
\textit{Zero-Shot Omni Models}
} \\
\multirow{2}{*}{
\shortstack[l]{MiniCPM-o 4.5}
}
& A
& 0.599
& \textbf{0.136}
& \textbf{0.509}
& \textbf{0.517} \\
& A-V
& \textbf{0.618}
& 0.110
& 0.474
& 0.496 \\
\addlinespace[1pt]
\multirow{2}{*}{
\shortstack[l]{Qwen2.5-Omni-7B}
}
& A
& \textbf{0.598}
& 0.135
& 0.496
& \textbf{0.557} \\
& A-V
& 0.535
& \textbf{0.151}
& \textbf{0.498}
& 0.491 \\
\multirow{2}{*}{
\shortstack[l]{Gemma-4-E4B-it}
}
& A
& 0.532
& \textbf{0.114}
& 0.474
& \textbf{0.509} \\
& A-V
& \textbf{0.558}
& 0.100
& \textbf{0.480}
& 0.475 \\
\multirow{2}{*}{
\shortstack[l]{Baichuan-Omni-1.5}
}
& A
& \textbf{0.497}
& \textbf{0.192}
& \textbf{0.516}
& \textbf{0.539} \\
& A-V
& 0.494
& 0.120
& 0.488
& 0.488 \\
\bottomrule
\end{tabular*}

\caption{
Input-modality ablation for binary, four-way, speech, and
environmental manipulation detection. A and A-V denote
audio-only and audio-visual inputs, respectively.
}
\label{tab:audio_visual_modality_comparison}
\end{table}

\paragraph{Setup.}
Unlike the preceding scene-consistency analysis, this
input-modality ablation examines the input requirements of the
manipulation-detection tasks introduced earlier. For the binary,
four-way, speech, and environmental targets, \emph{A} receives
only the final mixed audio, whereas \emph{A-V} receives the
identical audio together with its authentic video. Audio encoders and detectors
provide unimodal reference baselines. For each frozen A-V
encoder, the two input modes use independently fitted versions
of the same task-specific heads. Omni models receive the same
task prompt and audio in both modes, with video as the only
additional input. Results are averaged across the three core
protocols.

\paragraph{Broad acoustic pretraining provides strong unimodal baselines.}
Audio-only encoders and detectors perform strongly across all
four tasks. They capture diverse acoustic event patterns beyond
speech, supporting strong detection of both speech and
environmental manipulations. Among the pretrained audio
detectors, DF-Arena-1B and AudioMosaic achieve the strongest
overall results. This advantage is consistent with their
training coverage: DF-Arena-1B includes environmental deepfake
data alongside speech and singing corpora, while the evaluated
AudioMosaic checkpoint is fine-tuned on the EnvSDD TTA split.
These results highlight the value of combining broad acoustic
representations with direct exposure to environmental-sound
spoofing, rather than relying on speech anti-spoofing alone.

\paragraph{Frozen A-V encoders are consistently stronger with audio alone.}
For ImageBind, PE-AV, and CAV-MAE Sync, audio-only input
outperforms A-V input on every reported task. This consistent
pattern follows the benchmark target: the manipulation labels
are determined by the states of the audio components, while the
visual stream remains authentic across all classes. Audio
therefore provides the most direct forensic evidence for these
labels, whereas adding visual and relation features provides no
further separation under the unified lightweight-head protocol.
Although it does not improve manipulation classification, the
visual stream remains essential for scene-audio consistency,
providing the semantic reference needed to verify whether the
audio matches the visible scene.

\paragraph{Omni models show no systematic multimodal advantage.}
Adding video produces isolated binary improvements for some
omni models, but these gains do not extend consistently to
four-way or component-level detection. Audio-only input
therefore provides the more stable zero-shot setting for
forensic decisions.

\paragraph{Takeaway.}
Across the manipulation-detection tasks evaluated here, audio
provides the most reliable direct forensic signal. Broad audio
representations and domain-relevant anti-spoofing models provide
strong component cues, while frozen A-V encoders consistently
perform best in audio-only mode and omni models gain no
systematic benefit from video for these labels. The visual stream
remains central to modality-aware evaluation, as it anchors
scene-audio correspondence and enables models to be assessed
beyond acoustic artefact detection alone.